\documentclass[runningheads,a4paper]{llncs}

\usepackage{geometry}
\usepackage{amssymb}
\usepackage{graphicx}
\usepackage{url}
\usepackage{amsmath}

\begin{document}

\mainmatter 

\title{An overview of some semantic and syntactic \\ complexity classes}

\titlerunning{An overview of some semantic and syntactic complexity classes}

\author{ James L. Cox, \and Tayfun Pay}

\authorrunning{James L. Cox, Tayfun Pay}

\urldef{\mailsa}\path|tpay@gradcenter.cuny.edu|
\urldef{\mailsb}\path|cox@sci.brooklyn.cuny.edu|

\institute{{Brooklyn College \\ Computer and Information Science Department \\ 2900 Bedford Avenue,\\ Brooklyn, New York 11210\\ \mailsb}\\\ \and {Graduate Center of New York \\ Computer Science Department\\ 365 5$^{th}$ Avenue,\\ New York, New York 10016\\ \mailsa}}

\maketitle

\begin{abstract}
We review some semantic and syntactic complexity classes that were introduced to better understand the relationship between complexity classes {\bf P} and {\bf NP}. We also define several new complexity classes, some of which are associated with Mersenne numbers, and show their location in the complexity hierarchy.
\end{abstract}

\section{Introduction}
The study of non-deterministic polynomial-time Turing machines (NPTM) was initiated in \cite{C71} with the first NP-Complete problem. This problem asks if a Boolean expression in conjunctive normal form (CNF) has a satisfying truth assignment; and it is referred to as CNF-SAT. It was also shown that CNF-SAT is polynomial time Turing reducible to the problem of finding a tautology for a Boolean expression in disjunctive normal form (DNF), where this problem is CoNP-Complete. Then a whole array of other problems were shown to be NP-Complete in \cite{K72} via polynomial time many-one reductions to CNF satisfiability. The polynomial-time hierarchy, {\bf PH}, was introduced in \cite{S76} as a hierarchy of complexity classes that are derived from complexity classes {\bf NP} and {\bf CoNP} when used as oracles. A more elaborate list of NP-Complete problems were presented in \cite{GJ79}.

Researchers also examined versions of these NP-Complete problems with different constraints. For example, probabilistic polynomial time Turing machines and the corresponding PP-Complete problem Majority-SAT, which asks if a Boolean expression in CNF with $n$ variables have more than $2^{n-1}$ many satisfying truth assignments, was independently introduced in \cite{S75} and \cite{G77}. Complexity class {\bf C$_{=}$P} was also introduced in \cite{S75}, where the canonical {C$_{=}$P}-Complete problem asks if a Boolean expression in CNF with $n$ variables have exactly $2^{n-1}$ many satisfying truth assignments, and showed that {\bf C$_{=}$P} $\subseteq$ {\bf PP}. The complexity class {\bf \#P} was introduced in \cite{V79}, where it's complete problem \#SAT asks to find the total number of satisfying truth assignments for a Boolean expression in CNF. It was also proven in \cite{V79} that calculating the permanent of a 0-1 matrix is polynomial time Turing reducible to calculating the total number of accepting paths of a NPTM. It was shown in \cite{S77} that complexity classes {\bf \#P} and {\bf PP} are equivalent under polynomial time Turing reductions, that is {\bf P$^{\bf PP}$} = {\bf P$^{\bf \# P {[1]}}$}. Later on, the problems that ask if a Boolean expression in CNF has a unique satisfying truth assignment, and odd number of satisfying truth assignments were examined in \cite{BG82} and \cite{PZ83}, respectively. The complexity classes derived from these studies in \cite{BG82} and \cite{PZ83} are {\bf US} and {\bf$\oplus$P}, respectively. It was proven in \cite{T89} that {\bf PH} $\subseteq$ {\bf BPP$^{\bf \oplus P}$} $\subseteq$ {\bf P$^{\bf PP}$}, where {\bf BPP} is the bounded-error probabilistic polynomial time as defined in \cite{G77}. Some other intriguing complexity classes can be found in \cite{CH89} and \cite{FFK94}.

The study of categorical NPTM was initiated in \cite{V76} with the complexity class {\bf UP}. Other widely used terms instead of categorical are bounded, restricted and semantic. These complexity classes are not believed to possess complete problems, but they rather have promise problems. The promise problem for {\bf UP} is called Unambiguous-SAT and it asks the question that when promised to have a unique satisfying truth assignment or none, does the Boolean expression in CNF have a unique satisfying truth assignment? It was shown in \cite{VV86} that {\bf NP} $\subseteq$ {\bf RP}$^{Unambiguous-SAT}$, where {\bf RP} is the randomized polynomial time as defined in \cite{G77}. The semantic version of {\bf C$_{=}$P} was defined in \cite{BB92} as {\bf Half\_P}. Another notable semantic complexity class is {\bf EP}, which was defined in \cite{BHR00}. An {\bf EP} machine has an acceptance criterion of power of two and a rejection criterion of zero. It was shown in \cite{BHR00} that the syntactic version of {\bf EP}, called {\bf ES}, equals {\bf C$_{=}$P}. Various other interesting semantic complexity classes were introduced in \cite{AR88} and \cite{FFK94}.

In the next section, we review several reducibilities and establish what it means to be a semantic and syntactic complexity class. We also provide various examples of these complexity classes along with the proven relationships among them. In the section that follows, we define three new complexity classes, namely the semantic complexity classes {\bf MNP} and {\bf F$_=$P} and the syntactic complexity class {\bf MNS}. The complexity classes {\bf MNP} and {\bf MNS} are associated with Mersenne numbers whereas the complexity class {\bf F$_=$P} is closely related to the complexity class {\bf C$_=$P}. Then in the subsequent sections, we examine the relationship between these new complexity classes and already known ones. More precisely, we prove the following:

\textperiodcentered{ }{\bf FewP} $\subseteq$ {\bf MNP}

\textperiodcentered{ }{\bf US} $\subseteq$ {\bf MNS}

\textperiodcentered{ }{\bf PP} $\subseteq$ {\bf NP}$^{\bf MNS}$

\textperiodcentered{ }{\bf $\oplus$P} $\subseteq$ {\bf NP}$^{\bf MNS}$

\textperiodcentered{ }{\bf MNP} $\subseteq$ {\bf $\oplus$P}

\textperiodcentered{ }{\bf C$_=$P} $\subseteq$ {\bf MNS} and {\bf MNS} $\subseteq$ {\bf C$_=$P} which implies that {\bf MNS} = {\bf C$_=$P}

\textperiodcentered{ }{\bf MNP} $\cap$ {\bf EP} = {\bf UP}

\section{Definitions and Preliminaries}

We are interested in polynomial time Turing reducibility (also called Cook reducibility), polynomial time Post reducibility (also called truth table reducibility), polynomial time conjunctive and disjunctive truth table reducibilities and polynomial time many-one reducibility (also called Karp reducibility) as defined below.

\begin{definition}\normalfont

Let $A$ and $B$ be classes of languages.
\begin{enumerate}
\item $A$ is Turing reducible to $B$ in polynomial time, $(A_{\leq Tur}B)$, such that \\$A \in$ {\bf P}$^{B}$.

\item $A$ is Post reducible to $B$ in polynomial time, $(A_{\leq Post} B)$, such that $(\exists f \in {\bf FP})$ $(\exists C\in{\bf P})$ $(\forall x )$ $[(\exists c )(\exists y_{1}, y_{2}, \dots, y_{c})$ $[f(x)=y_{1}\#y_{2}\#\dots \#y_{c}\#]$ $\wedge$ $(x \in A \leftrightarrow x\#X_{B}(y_{1})X_{B}(y_{2}) \dots X_{B}(y_{c}) \in C)]$.

\item $A$ is conjunctive truth table reducible to $B$ in polynomial time, $(A_{\leq dtt} B)$, such that $(\exists f \in {\bf FP})$ $(\forall x )$ $[(\exists c )(\exists y_{1}, y_{2}, \dots, y_{C})$
$[f(x) = y_{1}\#y_{2}\#\dots \#y_{C}\#]$ $\wedge$ $( x \in A \leftrightarrow y_{1} \in B \wedge y_{2} \in B \wedge \dots \wedge y_{c} \in B)]$.

\item $A$ is disjunctive truth table reducible to $B$ in polynomial time, $(A_{\leq dtt} B)$, such that $(\exists f \in {\bf FP})$ $(\forall x )$ $[(\exists c )(\exists y_{1}, y_{2}, \dots, y_{C})$
$[f(x) = y_{1}\#y_{2}\#\dots \#y_{C}\#]$ $\wedge$ $( x \in A \leftrightarrow y_{1} \in B \vee y_{2} \in B \vee \dots \vee y_{c} \in B)]$.

\item $A$ is many-one reducible to $B$ in polynomial time, $(A_{\leq m} B)$, such that\\ $(\exists f \in {\bf FP})$ $(\forall x)[x \in A \leftrightarrow f(x) \in B]$.

\end{enumerate}

The following implications hold for all class of languages $A$ and $B$: \\

$(A_{\leq m} B)$ ${\genfrac{}{}{0pt}{}{\nearrow}{\searrow}}$ ${\genfrac{}{}{0pt}{}{(A_{\leq ctt} B)}{(A_{\leq dtt} B) }}$ ${\genfrac{}{}{0pt}{}{\searrow}{\nearrow}}$ $(A_{\leq Post} B)$ $\rightarrow$ $(A_{\leq Tur}B)$
\end{definition}

Another important reduction is the parsimonious reduction between functions that preserve the number of solutions. 

\begin{definition}\normalfont Let $F$ and $G$ be any functions. 
\begin{enumerate}
\item $F$ is parsimonious reducible to $G$ in polynomial time, $(F_{\leq par} G)$, such that $(\exists h \in  {\bf FP})$ $(\forall x)[ F(x) = G(h(x))]$, where $h$ is a total function. 
\end{enumerate}
\end{definition}

More detailed explanations about different types of reductions can be found in \cite{LLS75} and more peculiar ones are discussed in \cite{HO02}. It is well known that if a complexity class is closed under some reduction then a class reduced to it under that reduction is a subset of it.

One way to generalize complexity classes is through leaf languages, where language $L$ $\subseteq$ $\{0,1\}^{*}$. Then assume that leaf languages $L_{A}$ and $L_{R}$ have the property that $L_{A}$ $\cap$ $L_{R}$ = $\emptyset$, where $L_{A}$ is the acceptance criterion and $L_{R}$ is the rejectance criterion of a leaf language class. \cite{P94}

\begin{definition} \normalfont
A given complexity class is classified as a syntactic complexity class if and only if it has the property that $L_{A} \cup L_{R}$ = $\{0,1\}^{*}$.
\end{definition}

\begin{definition} \normalfont
A given complexity class is classified as a semantic complexity class if and only if it has the property that $L_{A} \cup L_{R}$ $\neq$ $\{0,1\}^{*}$.
\end{definition}

\begin{definition}\normalfont
A language $L$ is in semantic complexity class {\bf UP}, as defined in \cite{V76}, if there exist a polynomial $p$ and a polynomial-time predicate $R$ such that, for each $x$,

$x \in L \Rightarrow ||\{y| \ |y| = p(|x|) \wedge R(x, y)\}|| = 1$

$x \not \in L \Rightarrow ||\{y| \ |y| = p(|x|) \wedge R(x, y)\}|| = 0$
\end{definition}

\begin{definition}\normalfont
A language $L$ is in semantic complexity class {\bf UP}$_{O_{(k)}}$, as defined in \cite{B89}, if there exist a constant $k$ $>$ $1$, a polynomial $p$ and a polynomial-time predicate $R$ such that, for each $x$,

$x \in L \Rightarrow 1 \leq ||\{y| \ |y| = p(|x|) \wedge R(x, y)\}|| \leq k$

$x \not \in L \Rightarrow ||\{y| \ |y| = p(|x|) \wedge R(x, y)\}|| = 0$
\end{definition}

\begin{definition}\normalfont
A language $L$ is in semantic complexity class {\bf FewP}, as defined in \cite{A86}, if there exist polynomials $p$ and $q$ and a polynomial-time predicate $R$ such that, for each $x$,

$x \in L \Rightarrow 1 \leq ||\{y| \ |y| = p(|x|) \wedge R(x, y)\}|| \leq q(x) $

$x \not \in L \Rightarrow ||\{y| \ |y| = p(|x|) \wedge R(x, y)\}|| = 0$
\end{definition}

\begin{definition}\normalfont
A language $L$ is in semantic complexity class {\bf EP}, as defined in \cite{BHR00}, if there exist a polynomial $p$ and a polynomial-time predicate $R$ such that, for each $x$,

$x \in L \Rightarrow ||\{y| \ |y| = p(|x|) \wedge R(x, y)\}|| = 2^{t}$, where $t$ $\in \mathbb{N}_{0} = \{0,1,2, ...\}$


$x \not \in L \Rightarrow ||\{y| \ |y| = p(|x|) \wedge R(x, y)\}|| = 0$
\end{definition}

\begin{definition}\normalfont
A language $L$ is in semantic complexity class {\bf Half\_P }, as defined in \cite{BB92}, if there exist a polynomial $p$ and a polynomial-time predicate $R$ such that, for each $x$,

$x \in L \Rightarrow||\{y| \ |y| = p(|x|) \wedge R(x, y)\}|| = 2^{p(|x|)-1} $

$x \not \in L \Rightarrow ||\{y| \ |y| = p(|x|) \wedge R(x, y)\}|| = 0$
\end{definition}

It has been shown that {\bf P} $\subseteq$ {\bf UP} $\subseteq$ {\bf UP}$_{O_{(k)}}$ $\subseteq$ {\bf FewP} $\subseteq$ {\bf EP} $\subseteq$ {\bf NP} and that {\bf P} $\subseteq$ {\bf Half\_P} $\subseteq$ {\bf EP} $\subseteq$ {\bf NP}.

\begin{definition}\normalfont
A language $L$ is in syntactic complexity class {\rm \bf C$_{=}$P}, as defined in \cite{S75}, if there exist a polynomial $p$ and a polynomial-time predicate $R$ such that, for each $x$,

$x \in L \Leftrightarrow ||\{y| \ |y| = p(|x|) \wedge R(x, y)\}|| = 2^{p(|x|)-1} $
\end{definition}

An alternate definition of {\rm \bf C$_{=}$P} was provided in \cite{W86}, such that a language $L$ is in {\bf C$_=$P} if there exist a $f$ ${\in}$ {\bf FP} such that $x$ ${\in}$ $L$ if and only if the total number of accepting paths equals $f(x)$, for every $x$ ${\in}$ ${\Sigma^*}$.

\begin{definition}\normalfont
A language $L$ is in syntactic complexity class {\bf ES}, as defined in \cite{BHR00}, if there exist a polynomial $p$ and a polynomial-time predicate $R$ such that, for each $x$,

$x \in L \Leftrightarrow ||\{y| \ |y| = p(|x|) \wedge R(x, y)\}|| = 2^{t}$, where $t$ $\in \mathbb{N}_{0} = \{0,1,2, ...\}$
\end{definition}

\begin{definition}\normalfont
A language $L$ is in syntactic complexity class {\bf PP}, as defined in \cite{S75}, if there exist a polynomial $p$ and a polynomial-time predicate $R$ such that, for each $x$,

$x \in L \Leftrightarrow ||\{y| \ |y| = p(|x|) \wedge R(x, y)\}|| > 2^{p(|x|)-1} $

\end{definition}

\begin{definition}\normalfont
Functional complexity class {\bf \#P}, as defined in \cite{V79}, counts the total number of accepting paths of a NPTM.

{\bf \#P} = $\{f | (\exists$ a NPTM $M) (\forall x)$ $[f (x) = \#accept_{M}(x)]\}$.
\end{definition}

It was proven in \cite{BHR00,S75} that {\bf ES }$ = ${ \bf C$_{=}$P} $\subseteq$ {\bf PP} and in \cite{BBS86} that \\{\bf P}$^{\bf PP}$ = {\bf P}$^{{\bf\#P}[1]}$.

\begin{definition}\normalfont
A language $L$ is in syntactic complexity class {\bf US}, as defined in \cite{BG82}, if there exist a polynomial $p$ and a polynomial-time predicate $R$ such that, for each $x$,

$x \in L \Leftrightarrow ||\{y| \ |y| = p(|x|) \wedge R(x, y)\}|| = 1 $
\end{definition}

\begin{definition}\normalfont
A language $L$ is in syntactic complexity class {\bf $\oplus$P}, as defined in \cite{PZ83}, if there exist a polynomial $p$ and a polynomial-time predicate $R$ such that, for each $x$,

$x \in L \Leftrightarrow ||\{y| \ |y| = p(|x|) \wedge R(x, y)\}|| \not \equiv$ {\rm 0 (Mod 2)}
\end{definition}

\begin{definition}\normalfont
A language $L$ is in semantic complexity class {\rm \bf RP}, as defined in \cite{G77}, if there exist a polynomial $p$ and a polynomial-time predicate $R$ such that, for each $x$,

$x \in L \Rightarrow||\{y| \ |y| = p(|x|) \wedge R(x, y)\}|| \geq 2^{p(|x|)-1} +\epsilon $ where $ 0 < \epsilon < 2^{p(|x|)-1} $

$x \not \in L \Rightarrow ||\{y| \ |y| = p(|x|) \wedge R(x, y)\}|| = 0 $
\end{definition}

\begin{definition}\normalfont
A language $L$ is in semantic complexity class {\bf BPP}, as defined in \cite{G77}, if there exist a polynomial $p$ and a polynomial-time predicate $R$ such that, for each $x$,

$x \in L \Rightarrow ||\{y| \ |y| = p(|x|) \wedge R(x, y)\}|| \geq 2^{p(|x|)-1} +\epsilon $ where $ 0 < \epsilon < 2^{p(|x|)-1} $

$x \not \in L \Rightarrow ||\{y| \ |y| = p(|x|) \wedge R(x, y)\}|| \leq 2^{p(|x|)-1} +\epsilon $ where $ 0 < \epsilon < 2^{p(|x|)-1} $
\end{definition}

It was shown that {\bf CoNP} $\subseteq$ {\bf US} $\subseteq$ {\bf C$_{=}$P} and that {\bf Half\_P} $\subseteq$ {\bf RP} $\subseteq$ {\bf BPP} $\subseteq$ {\bf PP}. It has been proven in \cite{VV86} that {\bf NP} $\subseteq$ {\bf RP}$^{\bf \oplus P}$.

\begin{definition} \normalfont Polynomial Hierarchy, as defined in \cite{MS72}, for $i\geq0$ is

${\bf PH} = \bigcup_{i}\Sigma^{p}_{i}$,

$\Sigma^{p}_{i+1} = {{\bf NP}^{\Sigma^{p}_{i}}}$,

$\Pi^{p}_{i+1} = {{\bf CoNP}^{\Sigma^{p}_{i}}}$,

$\Delta^{p}_{i+1} = {{\bf P}^{\Sigma^{p}_{i}}}$,

where $\Sigma^{p}_{0} = \Pi^{p}_{0} = \Delta^{p}_{0} = \Delta^{p}_{1} = {\bf P},$ $\Sigma^{p}_{1} = {\bf NP},$ $\Pi^{p}_{1} = {\bf CoNP},$ and \newline $\Delta^{p}_{2} = {\bf P}^{{\bf NP}} = {\bf P}^{{\bf CoNP}} $.

\end{definition}

It has been proven in \cite{T89} that {\bf PH} $\subseteq$ {\bf BPP}$^{\bf \oplus P}$ $\subseteq$ {\bf P}$^{{\bf\#P}}$.

\begin{definition}\normalfont
Functional complexity class { \bf GapP}, as defined in \cite{FFK94}, is the difference between total number of accepting and rejecting paths of a NPTM.

{\bf GapP} = $\{f | (\exists$ a NPTM $M) (\forall x)$ $[f (x) = \#accept_{M}(x)-\#reject_{M}(x)]\}$.
\end{definition}

It can be easily seen that {\bf \#P} $\subseteq$ {\bf GapP}. Some complexity classes can also be defined using {\bf GapP} functions. For example, language $L$ is in {\bf C$_=$P} if there exist a $f$ ${\in}$ {\bf GapP} such that $x$ ${\in}$ $L$ if and only if ${f(x)=0}$ for every $x$ ${\in}$ ${\Sigma^*}$.

\begin{definition}\normalfont
A language $L$ is in semantic complexity class {\bf SPP}, as defined in \cite{FFK94}, if there exist a {\bf GapP} function $f$ such that, for each $x$

$x \in L \Rightarrow f(x) = 2$

$x \not \in L \Rightarrow f(x) = 0$
\end{definition}

The acceptance criterion is $2$ instead of $1$ because if the total number of computation paths are even then the difference between the total number of accepting and rejecting paths cannot be an odd number. {\bf SPP} is the smallest complexity class that can be defined using {\bf GapP} functions. It was shown in \cite{FFK94} that ${\bf UP} \subseteq {\bf SPP}$, ${\bf SPP} \subseteq {\bf \oplus P}$. It was also proven in \cite{FFK94} that ${\bf SPP}^{\bf SPP} = {\bf SPP}$, ${\bf C_=P}^{\bf SPP} = {\bf SPP}$ and ${\bf PP}^{\bf SPP} = {\bf PP}$. 

\begin{definition}\normalfont
A language $L$ is in semantic complexity class {\bf WPP}, as defined in \cite{FFK94}, if there exist a {\bf GapP} function $f$ and a {\bf FP} function g such that, for each $x$

$x \in L \Rightarrow f(x) = g(x)$

$x \not \in L \Rightarrow f(x) = 0$
\end{definition}

It was shown in \cite{FFK94} that ${\bf SPP} \subseteq {\bf WPP} \subseteq {\bf C_=P}$.

\section{Even more complexity classes}
\subsection{Mersenne Number Satisfiability}
Mersenne numbers are named after Marin Mersenne whom began the study of these numbers in the 17$^{th}$ century. A Mersenne number is a positive integer that is one less than a power of two,
$M_{n} = 2 ^{n}-1$ and consists of all 1 bits in its binary representation. It is well known that if $M_{n}$ is a prime number then $n$ is a prime number as well. The study of Mersenne primes has been an alluring field with the emergence of powerful computers, which can do calculations that would be very-hard for humans to do by hand. There have been 50 Mersenne primes discovered to date and the largest known Mersenne prime is $2^{77,232,917}-1$.

\begin{definition}\normalfont A language $L$ is in syntactic complexity class {\bf MNS} if there exist polynomial $p$ and a polynomial-time predicate $R$ such that, for each $x$,

$x \in L \Leftrightarrow ||\{y| \ |y| = p(|x|) \wedge R(x, y)\}|| = 2^{t}-1$, where $t$ $\in \mathbb{N}_{>0} = \{1,2,3, ...\}$
\end{definition}

\begin{definition}\normalfont Mersenne-Number-SAT

Given a Boolean expression in CNF, is it true that it has Mersenne number of satisfying truth assignments?
\end{definition}

\begin{theorem}\normalfont Mersenne-Number-SAT is {\bf MNS}-complete
\begin{proof}\normalfont
Mersenne-Number-SAT is clearly in {\bf MNS} given the above definition. Completeness follows from the parsimonious reductions in \cite{V79} of any problem in {\bf \#P} to \#SAT.$\qed$
\end{proof}
\end{theorem}

\begin{definition}\normalfont A language $L$ is in semantic complexity class {\bf MNP} if there exist polynomial $p$ and a polynomial-time predicate $R$ such that, for each $x$,

$x \in L \Rightarrow ||\{y| \ |y| = p(|x|) \wedge R(x, y)\}|| = 2^{t}-1$, where $t$ $\in \mathbb{N}_{>0} = \{1,2,3, ...\}$

$x\not \in L \Rightarrow ||\{y| \ |y| = p(|x|) \wedge R(x, y)\}|| = 0 $

\end{definition}

\begin{definition}\normalfont Promise-Mersenne-Number-SAT

Given a Boolean expression in CNF that is promised to have Mersenne number of satisfying truth assignments or none, does it have Mersenne number of satisfying truth assignments?
\end{definition}

We obtain the following inclusions just from the above definitions:

\textperiodcentered{ }{\bf MNP} $\subseteq$ {\bf MNS}

\textperiodcentered{ }{\bf MNP} $\subseteq$ {\bf NP}

\textperiodcentered{ }{\bf MNS} $\subseteq$ {\bf P$^{\bf PP}$} = {\bf P$^{\bf \# P {[1]}}$}

\subsection{A long lost semantic relative of ${\bf C_=P}$}

A perceptive reader would have noted that two out of the three alternate definitions of ${\bf C_=P}$
that employ {\bf \#P} functions have their own semantic versions, namely {\bf Half\_P} and {\bf EP}. And by definition ${\bf Half\_P} \subseteq {\bf EP}$. It is not known whether they are equal, although their syntactic versions have been proven to be equal in \cite{BHR00}. We next define the semantic version of the third alternate definition of ${\bf C_=P}$ that also employ {\bf \#P} functions.

\begin{definition}\normalfont
A language $L$ is in semantic complexity class {\bf F$_{=}$P} if there exist a polynomial $p$, a polynomial-time predicate $R$ and a polynomial time computable function $f$, such that, for each $x$,

$x \in L \Rightarrow ||\{y| \ |y| = p(|x|) \wedge R(x, y)\}|| = f(x)$

$x \not \in L \Rightarrow ||\{y| \ |y| = p(|x|) \wedge R(x, y)\}|| = 0$
\end{definition}

It is not hard to see that we would have still obtained the same complexity class if we had changed the {\bf GapP} function in the definition of {\bf WPP} to a {\bf \# P} function.

\begin{definition}\normalfont Promise-Exact-Number-SAT

Given a Boolean expression in CNF that is promised to have $f(x)$ many satisfying truth assignments or none, does it have $f(x)$ many satisfying truth assignments?
\end{definition}

We obtain the following inclusions just from the above definitions:

\textperiodcentered{ }{\bf FewP} $\subseteq$ {\bf F$_{=}$P}

\textperiodcentered{ }{\bf F$_{=}$P} $\subseteq$ {\bf C$_{=}$P}

\textperiodcentered{ }{\bf F$_{=}$P} $\subseteq$ {\bf WPP}

\section{Relationship with complexity class {\bf FewP}}

\begin{definition}\normalfont Non-gappy \cite{BHR00}

\normalfont Let $S$ be any set of positive integers. Then $S$ is non-gappy if $S \not = 0$ and $(\exists k > 0)(\forall n \in S)(\exists m \in S)[m > n \wedge m/n \leq k]$.
\end{definition}

\begin{definition}\normalfont P-printable \cite{JY84}

\normalfont Let $L$ be any subset of $\Sigma ^* $. Then $L$ is P-printable if there is a deterministic Turing machine $M$ that runs in polynomial-time such that, for every nonnegative integer $n$, $M(0^n)$ prints out the set $\{ x | x
\in L \wedge |x| \leq n\}$.
\end{definition}

Furthermore, Theorem 3.4 in \cite{BHR00} states that, ``Let T be any set of positive integers such that T has a non-gappy, P-printable subset. Then {\bf FewP} is contained in any complexity class with the acceptance criterion of T and rejectance criterion of zero."

\begin{theorem} \normalfont {\bf FewP} $\subseteq$ {\bf MNP}
\begin{proof} \normalfont

The acceptance criterion of {\bf MNP} is clearly Non-gappy and P-printable according to the above definitions. Then given Theorem 3.4 in \cite{BHR00}, {\bf FewP} is contained in {\bf MNP}.

Therefore, {\bf FewP} $\subseteq$ {\bf MNP}.$\qed$
\end{proof}
\end{theorem}

\section{Relationship with complexity class {\bf US}}

\begin{theorem} \normalfont {\bf US} $\subseteq$ {\bf MNS}
\begin{proof} \normalfont Suppose we have a {\bf US} machine $M$. We construct machine $M'$, that originally has $2^n$ many paths. Then on each one of it's accepting paths, machine $M'$ non-deterministically decides to accept on $2^n-1$ many paths. Then we claim that machine $M'$ will have a Mersenne number of accepting paths if and only if machine $M$ has a single accepting path. To observe that this is true, note that if the number of accepting paths of the original machine $M$ is $k$, then $k$ must satisfy $k (2^n-1) = 2^m-1$, where $m$ is a positive integer. Thus $k = \dfrac{2^m-1}{ 2^n-1 }$. The general solution of this equation for $k \in \mathbb{N}_{>0}$ is $m = \dfrac {2 i \pi k}{\ln{2}}$, where $i$ is an imaginary number. The first two values of $k$ that make $m$ an integer (and in fact a real number) are $1$ and $2^n+1$. Since $k \in \{1,...,2^n\}$, this means that $M'$ will have a Mersenne number of accepting paths if and only if $k=1$. As a result, machine $M'$ is a {\bf MNS} machine.

Therefore, {\bf US} $\subseteq$ {\bf MNS}.$\qed$ \end{proof}
\end{theorem}

We should note that there are two well known CoNP-complete problems, namely the CNF contradiction and the DNF tautology, where the former asks if no truth assignment satisfies a Boolean expression in CNF and the latter asks if all possible truth assignments satisfies a Boolean expression in DNF. When one was shown to be CoNP-complete, the other was easily shown to be CoNP-complete by simply reversing the accepting and rejecting states of a NPTM. Then by employing the same methodology, we can easily show that Maximum-Mersenne-Number SAT, which asks if a Boolean expression in DNF with $n$ variables has $2^n-1$ many satisfying truth assignments, is US-complete. Note that finding a unique satisfying truth assignment for a Boolean expression in DNF and finding a maximum Mersenne number of satisfying truth assignments for a Boolean expression in CNF are both computable in polynomial time.

\section{Relationship with complexity class {\bf PP}}

In order to prove the following result using a NPTM equipped with a {\bf MNS} oracle to recognize Majority-SAT, we must overcome a technical difficulty that occurs if the input formula is unsatisfiable. We finesse this problem by initially running a standard NPTM for SAT. And we only begin a new simulation regime and make queries on accepting paths. In this way, we will never make queries if the input is unsatisfiable and we will reject outright.

\begin{theorem} \normalfont {\bf PP} $\subseteq$ {\bf NP$^{\bf MNS}$}

\begin{proof} \normalfont We show that a {\bf NP} machine with access to a {\bf MNS} oracle solves the PP-complete problem Majority-SAT.

Assume that we have an {\bf NP} machine $M$ so that the number of satisfying truth assignments to Boolean expression $x$ is equal to the number of accepting paths of $M$ on input $x$. Also, assume that $x$ has $n$ variables and thus $M(x)$ has $2^{n}$ many paths. Then we construct an {\bf NP} machine $M'$ with access to oracle {\bf MNS}. Initially, machine $M'$ simulates machine $M$ on input $x$. If $x$ is unsatisfiable then all paths of $M'$ will reject. On the other hand, if $x$ has at least one satisfying truth assignment then $M'$ will reach a state where $M$ would have accepted. At this point, $M'$ enters a query state to determine if $M$ would have accepted on more than half of the paths. Then $M'$ non-deterministically selects a number $k$ from $0$ to $2^{n-1}-1$. After that, for each choice of $k$, we construct another machine $M''$ that does the following. It non-deterministically chooses to accept on $2^{n} -1 + k$ many paths and simulates $M$ on the other paths. Then we query the {\bf MNS} oracle with the Boolean expression $f(M'',x)$ and $M'$ accepts if the query answers $YES$.

Observation 1: The fact that we entered the query states implies that the Boolean expression $x$ has at least one satisfying truth assignment.

Observation 2: Let $M$ have $p$ many accepting paths on input $x$. Then $M''$ has at most $2^{n} - 1 + k + p$ many accepting paths by construction. Since $p>0$ by construction as well, then $M''$ has at least $2^{n}$ many accepting paths.

Observation 3: $M''$ has less than $2^{n+2}-1$ many accepting paths for any choice of $k$. The maximum number of accepting paths that $M''$ can have is when $k = 2^{n-1}-1$ and the original input formula is a tautology. This results in $(2^{n} -1) + (2^{n-1}-1) + (2^{n})$ many accepting paths, which equals $2^{n+1} + 2^{n-1}-2$ and is clearly less than $2^{n+2}-1$.

Observation 4: If $p \leq 2^{n-1}$ then $M''$ accepts on at most $2^{n+1} -2$ paths. If $p > 2^{n-1}$ then for some choice of $k$, $M''$ will accept on exactly $2^{n+1} -1$ many paths.

From the observations above, the number of accepting paths of $M''$ lies between $2^{n} -1$ and $2^{n+2}-1$. The only Mersenne number of accepting paths that $M''$ can have is $2^{n+1} -1$, which is achieved when $p > 2^{n-1}$. Thus if one of these queries gives the answer $YES$ then $x$ is in Majority-SAT.

Therefore, {\bf PP} $\subseteq$ {\bf NP$^{\bf MNS}$}.$\qed$\end{proof}

\end{theorem}

\begin{lemma} \normalfont {\bf P$^{\bf PP}$} $\subseteq$ {\bf P$^{\bf NP^{\bf MNS}}$}

It follows from Theorem 4.
\end{lemma}

\begin{lemma} \normalfont {\bf PH} $\subseteq$ {\bf P$^{\bf NP^{\bf MNS}}$}

It follows from Theorem 4 and Toda's Theorem \cite{T89}, {\bf PH} $\subseteq$ {\bf P$^{\bf PP}$}.
\end{lemma}

\section{Relationship with complexity class {\bf $\oplus$P} }

\begin{theorem} \normalfont {\bf $\oplus$P} $\subseteq$ {\bf NP$^{\bf MNS}$}
\begin{proof} \normalfont We show that a {\bf NP} machine with access to a {\bf MNS} oracle solves the $\oplus$P-complete problem Parity-SAT.

Assume that we have an {\bf NP} machine $M$ so that the number of satisfying truth assignments to the Boolean expression $x$ is equal to the number of accepting paths of $M$ on input $x$. Also, assume that $x$ has $n$ variables and thus $M(x)$ has $2^{n}$ many paths. Then we construct an {\bf NP} machine $M'$ with access to oracle {\bf MNS} that on input $x$ behaves as follows: It first non-deterministically selects an even number $k$, where $0 \leq k < 2^n$. Then we construct a machine $M''$ that does the following. It non-deterministically chooses to accept on $k$ paths and simulates $M$ on the other path. Then we query the {\bf MNS} oracle with the Boolean expression $f(M'',x)$ and $M'$ accepts if the query answers $YES$.

If $M$ accepts on an even number of paths, then clearly all queries answer no and $M'$ rejects. If $M$ accepts on some odd number of paths, say $j$, then there exists an even $k$, where $0 \leq k < 2^n$, such that $k+j = (2^i -1)$ for some $i$, and thus $M''$ accepts on Mersenne number of paths. As a result, machine $M'$ will accept on this query.

Therefore, {\bf $\oplus$P} $\subseteq$ {\bf NP$^{\bf MNS}$}.$\qed$\end{proof}
\end{theorem}

\begin{lemma}\normalfont {\bf BPP$^{\bf \oplus P}$} $\subseteq$ {\bf BPP$^{\bf NP^{\bf MNS}}$}

It follows from Theorem 5.

\end{lemma}

\begin{lemma} \normalfont {\bf PH} $\subseteq$ {\bf BPP$^{\bf NP^{\bf MNS}}$}

It follows from Theorem 5 and Toda's Theorem \cite{T89}, which states that \\ {\bf PH} $\subseteq$ {\bf BPP$^{\bf \oplus P}$}.

\end{lemma}

However, clearly Lemma 2 is a stronger result than Lemma 4 since \\ {\bf P$^{\bf NP^{\bf MNS}}$} $\subseteq$ {\bf BPP$^{\bf NP^{\bf MNS}}$}

\begin{theorem} \normalfont {\bf MNP} $\subseteq$ {\bf $\oplus$P}
\begin{proof}\normalfont
We are given a machine $M$ that decides some {\bf MNP} promise problem $pp$. If input $x$ should be accepted, then machine $M$ accepts on a Mersenne number of paths. If input $x$ should be rejected, then machine $M$ has no accepting paths. Clearly, machine $M$ satisfies the criterion for a {\bf $\oplus$P} machine, since all Mersenne numbers are odd. Therefore, promise problem $pp$ is in {\bf $\oplus$P}.$\qed$\end{proof}
\end{theorem}

\section{Relationship with complexity class {\bf C$_{=}$P} }

It is easy to see that we can revise Theorem 4 and achieve {\bf C$_{=}$P} $\subseteq$ {\bf NP$^{\bf MNS}$}. However, we can actually do much better.

\begin{theorem}\normalfont {\bf C$_{=}$P} $\subseteq$ {\bf MNS}

\begin{proof} \normalfont Suppose we have a {\bf C$_{=}$P} machine $M$. We design a {\bf MNS} machine $M'$ that accepts the same language as $M$. Assume for ease of presentation that on input $x$ our machine $M$ has $2^{n+1}$ many total paths and thus accepts $x$ if and only if it accepts on exactly $2^{n}$ many paths. We also assume without loss of generality that $n$ is sufficiently large. Then machine $M'$ on input $x$ immediately accepts on $2^{2n}-1$ many paths and simulates $M$ the following way: 1) For each of $M$'s rejecting states, it accepts and 2) For each of $M$'s accepting states, it accepts on 2$^{n}-1$ many paths. We claim that $M'$ accepts if and only if $M$ accepts on $2^{n}$ many paths. We show this with the following three lemmas.

\begin{lemma} \normalfont If $M$ accepts on $2^{n}$ many paths, then $M'$ accepts on Mersenne number of paths.

If $M$ accepts on $2^{n}$ many paths then $M'$ accepts on $(2^{2n}-1) + (2^{n}*(2^{n}-1)) + (2^{n})$ many paths. This expression evaluates to $2^{2n} -1 + 2^{2n} - 2^{n} + 2^{n}$, which simplifies to $2^{2n +1}-1$. Therefore, $M'$ will accept since $2^{2n +1}-1$ is a Mersenne number.
\end{lemma}

\begin{lemma} \normalfont If $M$ accepts on less than $2^{n}$ many paths, then $M'$ does not accept on Mersenne number of paths.

Machine $M'$ will accept on greater than $2^{2n}-1$ many paths by construction, since it also accepts on rejecting paths. Furthermore, $M'$ will accept on less than $2^{2n +1}- 1$ many paths once again by construction. Thus the number of accepting paths of $M'$ will fall in between two consecutive Mersenne numbers. Therefore, $M'$ will reject.

\end{lemma}

\begin{lemma}\normalfont If $M$ accepts on more than $2^{n}$ many paths then $M'$ does not accept on Mersenne number of paths.

Machine $M'$ will accept on greater than $2^{2n+1}-1$ many paths by construction. Then the maximum number of accepting paths is achieved for $M'$ when $M$ accepts on $2^{n+1}$ many paths. In this case, $M'$ will accept on $(2^{2n}-1) + ((2^{n+1}) * (2^{n}-1))$ many paths once again by construction. This expression equals $2^{2n+1} + 2^{2n} - 2^{n+1} - 1$ which is less than $2^{2n +2} - 1$. Thus the number of accepting paths of $M'$ will fall in between two consecutive Mersenne numbers. Therefore, $M'$ will reject.

\end{lemma}
As shown by the previous three lemmas that the {\bf MNS} machine $M'$ accepts if and only if the {\bf {C$_{=}$P}} machine $M$ accepts.

Therefore, {\bf C$_{=}$P} $\subseteq$ {\bf MNS}.$\qed$ \end{proof}

\end{theorem}

\begin{theorem} \normalfont {\bf MNS} $\subseteq$ {\bf C$_{=}$P}
\begin{proof}
\normalfont We provide a polynomial time disjunctive truth-table reduction from Mersenne-number-SAT to a language in {\bf C$_{=}$P}. It is well-known that {\bf C$_{=}$P} is closed under polynomial-time disjunctive truth table reductions, which was proven in \cite{BC93}. Assume that we have a {\bf MNS} machine $M$ that recognizes Mersenne-number-SAT. Let $x$ be an input with $n$ variables. We next show the disjunctive truth-table reduction to the canonical C$_{=}$P-complete problem Equal-SAT. Assume that we have a polynomial time machine $M'$ that carries out the reduction. For each $i \in \{0,...,n\}$, machine $M'$ constructs a machine $M_i$ that does the following. It immediately accepts on $2^n-2^i$ many paths and rejects on $2^i$ many paths, and also simulates machine $M$. Then for each $i$, we query the Equal-SAT oracle with the Boolean expression $f(M_i,x)$ and $M'$ accepts if any of these queries answer $YES$.

Therefore, {\bf MNS} $\subseteq$ {\bf C$_{=}$P} .$\qed$ \end{proof}

\end{theorem}

\begin{corollary} \normalfont{\bf MNS} $=$ {\bf C$_{=}$P}

Immediate consequence of Theorems 7 and 8.
\end{corollary}

\section{Conclusion}

We introduced two new semantic complexity classes and one new syntactic complexity class; and showed their location in the complexity hierarchy. It ended up being the case that {\bf MNS} actually equals {\bf C$_{=}$P}. However, a simple padding argument would have not yielded the result in Theorem 7, that is {\bf C$_{=}$P} $\subseteq$ {\bf MNS}. What our proof of Theorem 7 actually demonstrates is that given a Boolean expression $F$ in CNF with $n$ variables, one can in polynomial time construct $F'$ with $m$ variables so that the number of satisfying truth assignments to $F'$ is guaranteed to lie between $2^{m-2}-1$ and $2^{m}-1$. In fact, $F'$ will have exactly $2^{m-1}-1$ satisfying truth assignments if and only if $F$ was satisfied by exactly half of its assignments.

On the other hand, the relationship between {\bf EP} and {\bf MNP} is not so clear other than the fact that their intersection equals {\bf UP}, that is {\bf EP} $\cap$ {\bf MNP} = {\bf UP}. Although, we can change the acceptance criterion of a {\bf MNS} machine to be some specific power of two using the methodology in Theorem 7, we cannot necessarily do the same thing with the acceptance criterion of a {\bf MNP} machine. This is because we would also need to consider the rejection criterion of a {\bf MNP} machine and the result in Theorem 7 does not yield zero accepting paths when the original machine has zero accepting paths. However, {\bf MNP} can be viewed as the analog of {\bf EP} that is contained in {\bf $\oplus$P}, which {\bf EP} is not known to be. In fact, a relativized world was shown in \cite{BHR00} such that $\exists {\bf A }, {\bf EP ^{\bf A}} \not \subseteq {\bf \oplus P ^{\bf A}}$.

Another interesting question arises with respect to the relationship between {\bf F$_=$P}, and {\bf EP} and {\bf MNP}. There does not seem to be a straightforward proof to show any type of inclusion among them. However, it seems more likely that both {\bf EP} and {\bf MNP} are contained in {\bf F$_=$P}. Also, is {\bf F$_=$P} contained in {\bf $\oplus$P} just like {\bf MNP}; or does there exists a relativized world where {\bf F$_=$P} is not contained in {\bf $\oplus$P}, just like {\bf EP}.

Finally, we would like to mention that we attempted to change the base machine in Theorem 5 from {\bf NP} to {\bf RP} with no success. We also tried to derive a result just like Theorem 7 between {\bf $\oplus$P} and {\bf MNS}, but once again we were not able to do better than {\bf NP$^{\bf MNS}$}. However, we do think strongly about the possibility of {\bf $\oplus$P} $\subseteq$ {\bf RP$^{\bf MNS}$} or even {\bf $\oplus$P} $\subseteq$ {\bf MNS}.

\end{document}